\documentclass[twocolumn]{aastex631}

\usepackage{graphicx}	
\usepackage{amsmath}	
\usepackage{amssymb}	
\usepackage{amssymb}
\usepackage{upgreek}
\usepackage{bbm}
\usepackage{dsfont}
\usepackage{sansmath}
\usepackage{mathrsfs}
\usepackage{yfonts}
\usepackage{euscript}

\def\d{{\rm d}}
\def\D{{\rm D}}
\def\del{\partial}
\def\bfnabla{\mbox{\boldmath{$\nabla$}}}
\def\grad-s{\mbox{\boldmath{$\nabla$}}_{\!\! s}\,}
\def\bfx{{\bf x}}

\def\bfv{{\bf v}}
\def\bfk{{\bf k}}
\def\bfn{{\bf n}}

\def\gsim{\;\rlap{\lower 2.5pt\hbox{$\sim$}}\raise 1.5pt\hbox{$>$}\;}
\def\lsim{\;\rlap{\lower 2.5pt\hbox{$\sim$}}\raise 1.5pt\hbox{$<$}\;}

\def\gsim{\;\rlap{\lower 2.5pt\hbox{$\sim$}}\raise 1.5pt\hbox{$>$}\;}
\def\lsim{\;\rlap{\lower 2.5pt\hbox{$\sim$}}\raise 1.5pt\hbox{$<$}\;}
\def\del{{\partial}}
\def\grad{\mbox{\boldmath{$\nabla$}}}

\def\beq{\begin{equation}}
\def\eeq{\end{equation}}

\usepackage{color}
\definecolor{com_red}{rgb}{.8,.2,0.1}
\definecolor{com_turq}{rgb}{.0,.4,0.6}
\definecolor{ins_green}{rgb}{.1,.5,0.1}

\shorttitle{Early-time, small-scale structures in hot-exoplanet 
atmosphere simulations}
\shortauthors{Skinner \& Cho}
\graphicspath{./figs/}

\begin{document}

\title{Early-time small-scale structures in hot-exoplanet 
atmosphere simulations}

\author[0000-0002-5263-385X]{J. W. Skinner $^{\dagger,} $}
\affiliation{Division of Geological and Planetary Sciences, 
California Institute of Technology, 1200 E California Blvd, 
Pasadena, CA 91125, USA}
\affiliation{Martin A. Fisher School of Physics, Brandeis 
University, 415 South Street, Waltham, MA 02453, USA}
\email{$\dagger \,$ jskinner@caltech.edu}

\author[0000-0002-4525-5651]{J.Y-K. Cho}
\affiliation{Martin A. Fisher School of Physics, Brandeis 
University, 415 South Street, Waltham, MA 02453, USA}
 
\begin{abstract}
We report on the critical influence of small-scale flow 
structures (e.g., fronts, vortices, and waves) that 
immediately arise in hot-exoplanet atmosphere simulations 
initialized with a resting state.   
A hot, 1:1~spin--orbit synchronized Jupiter is used here 
as a clear example; but, the phenomenon is generic and 
important for any type of a hot synchronized 
planet---gaseous, oceanic, or telluric.  
When the early-time structures are not captured in 
simulations (due to, e.g., poor resolution and/or too 
much dissipation), the flow behavior is markedly different 
at later times---in an observationally significant way; 
for example, the flow at large-scale is smoother and much 
less dynamic.
This results in the temperature field, and its 
corresponding thermal flux, to be incorrectly predicted 
in numerical simulations, even when the quantities are 
spatially averaged.
\end{abstract}


\section{Introduction} \label{sec:intro}

A hallmark of nonlinear dynamical systems is their sensitivity to 
initial condition \citep{Poincare1914,Lorenz63}.  
In such systems, infinitesimal perturbations at early times are 
quickly amplified by the evolution, leading to loss of 
predictability in certain variables and measures \citep{Lorenz64}. 
This phenomenon, often referred to as the ``butterfly 
effect'', lies at the heart of chaos and underscores the inherent
unpredictability of many complex systems.
As a highly nonlinear system, the hot-exoplanet atmosphere also 
exhibits this paradigmatic feature.  
Indeed, numerical simulations of hot-exoplanet atmospheres are 
sensitive to their initial states and their ability to represent 
the flows across an adequate range of dynamically significant 
scales \citep[e.g.,][]{Choetal08,ThraCho10,Choetal15,SkinCho20}. 

Early efforts to simulate hot, 1:1 spin--orbit synchronized 
exoplanet atmospheres have utilized a very simple setup for 
the initial and forcing conditions \citep[e.g.,][]{Showman02,
  Choetal03,Cooper06,Choetal08,Dobbs-Dixon_2008,Showmanetal08a,
  Menou09,RausMen10}.
In this setup, an atmosphere initially at rest is set in motion 
by ``relaxing'' the temperature field to a prescribed spatial
distribution, on a specified timescale.
Despite its simplicity, the setup is useful and provides 
valuable insights---especially in the absence of detailed 
information about the atmosphere.
More sophisticated thermal forcing treatments include 
arbitrary choices and simplifications of complex (and often 
poorly-known) processes, which can hamper accuracy as well as 
understanding.
For this reason, simple setups continue to be utilized in
modeling studies today \citep[e.g.,][]{Debras2020, boning2024}.

A salient feature that can be studied with the simple setup 
for hot-exoplanets is the wide range of thermal relaxation 
timescales in the currently observable region of their 
atmospheres.
In particular, on hot-exoplanets the timescale can be very 
short (i.e., much shorter than the advective time scale) 
above the $\sim$$10^5$\,Pa altitude level. 
However, it has long been known that such quick adjustments 
from rest leads to a flow which is very sensitive to a wide 
range of modeling parameters 
\citep[e.g.,][]{Choetal03,ThraCho10,ThraCho11,PoliCho12,
Choetal15}. 
For example, numerical resolution, initial flow state, 
thermal relaxation timescale, strength and form of 
numerical dissipation, and altitude of peak heat absorption 
all affect the predictions \citep[e.g.,][]{Choetal03,Choetal08,
  Heng2011,Polietal14,SkinCho20,HammondAbbot22,Skinetal23}. 
 
In this paper, we highlight the profound effect of {\it small-scale 
structures that arise at early-times} on the late-time flow. 
This ``early-time sensitivity'' has not been explicitly called 
to attention before.
Past studies have generally focused on the state of the flow a long 
time after the start of the simulation (often referred to as the 
``equilibrated'' state\footnote{Presently, there is no universally 
accepted unique state or an ``equilibration time'' for hot-Jupiters, 
as both depend on the physical setup, initial condition, and 
numerical algorithm of the simulation \citep[see, e.g.,][]{Choetal08,
  ThraCho10,Choetal15,SkinCho20,Skinetal23}.}), generally 
employing only low to moderate resolution.
Little focus has been given to the transient, small-scale flow 
structures that occur during the first $\sim$10~days of the 
simulation. 
The tacit assumption in the past has been that the flow evolution 
would ``forget'' the initial condition and head inexorably to the 
same statistically-steady state.
Here simulations are performed at high resolution with low 
dissipation (to be elucidated below) to more accurately capture 
the small-scale dynamics than have been in the past. 

\section{Model}\label{sec:model}

The governing equations, planetary parameters, numerical 
model, and physical setup in this work are same as those 
in \cite{Choetal21} and \cite{SkinCho21}. 
Therefore, only a brief summary is presented here; 
we refer the reader to the above works for more 
details---as well as to \cite{SkinCho20}, \cite{Polietal14}, 
and \cite{Choetal15} for extensive convergence tests and 
inter-model comparisons. 
As in all of the aforementioned works, the hydrostatic 
primitive equations~(PE) are solved here to study the 
three-dimensional~(3D) atmospheric dynamics.
The dissipative PE, with pressure $p$ serving as the 
vertical coordinate, read:
\begin{subequations}\label{eq:PE}
 \begin{eqnarray}
    \frac{\partial{\zeta}}{\partial t}\ \ & = &\  
     \bfk \cdot \bfnabla\times \bfn\ +\ {\cal D}_\zeta \\
    \frac{\partial{\delta}}{\partial t}\ \ & = &\ 
     \bfnabla \cdot \bfn\ -\ \nabla^2\left(\frac{1}{2}\bfv^2\, +\, 
     \Phi\right)\ +\ {\cal D}_\delta \\
       \frac{\partial{\Theta}}{\partial t}\ & = &\ 
     -\bfnabla \cdot \Big(\Theta\, \bfv\Big)\ -\ 
     \frac{\del\ }{\del p}\Big(\omega\,\Theta\Big)\ +\ 
     \frac{\dot{\cal Q}}{\Pi}\,\ +\ 
     {\cal D}_\Theta \qquad \quad \\
    \frac{\partial \Phi}{\partial \Pi}\ & = &\ -\Theta \\
    \frac{\partial \omega}{\partial p}\ & = &\  -\delta\ . 
 \end{eqnarray}
\end{subequations}

In equations~(\ref{eq:PE}), 
$\zeta(\bfx,t) \equiv \bfk\!\cdot\!\bfnabla\!\times\!\bfv$ 
is the vorticity and 
$\delta(\bfx,t) \equiv \bfnabla\!\cdot\!\bfv$ is the 
divergence, where $\bfx \in \mathbb{R}^3$, $\bfk$ is the 
local vertical direction, $\bfv(\bfx,t)$ is the horizontal 
velocity, and $\bfnabla$ is the gradient which operates 
along an isobaric (constant $p$) surface; 
$\Theta(\bfx,t) \equiv (c_p/\Pi)\, T$ is the potential 
temperature, where 
$c_p$ is the constant specific heat at constant~$p$, 
$\Pi \equiv c_p (p/p_\text{\tiny ref})^\kappa$ is the 
Exner function with $\kappa \equiv {\cal R}/c_p$, 
${\cal R}$ the specific gas constant, and 
$p_\text{\tiny ref}$ a reference $p$, and
$T(\bfx,t)$ is the temperature; 
$\Phi(\bfx,t) \equiv gz$ is the specific geopotential, 
where $g$ is a constant and $z(\bfx,t)$ is the height; 
$\omega \equiv \D p / \D t$ is the vertical velocity, 
where 
$\D / \D t \equiv \del / \del t + \bfv\!\cdot\! \bfnabla 
+ \omega\,\del / \del p\,$ is the material derivative;
$\mathbf{n} \equiv -( \zeta + f)\,\bfk\times\bfv - 
\delta\bfv - \del(\omega \bfv) / \del p$, where 
$f \equiv 2\Omega\sin\phi$ is the Coriolis parameter with 
$\Omega$ the planetary rotation rate and $\phi$ the latitude; 
$\rho = p/({\cal R}\,T)$ is the density; 
${\cal D}_\zeta(\nu,\mathfrak{p})$, 
${\cal D}_\delta(\nu,\mathfrak{p})$, and 
${\cal D}_\Theta(\nu,\mathfrak{p})$ are the 
(hyper)dissipations, which are dependent on the dissipation
coefficient $\nu = \nu(\mathfrak{p})$ and order 
$\mathfrak{p} \in \mathbb{Z}^+$; and, 
$\dot{\cal Q}(\bfx,t;\tau_r)$ 
is the net heating rate, where $\tau_r$ is the relaxation 
time parameter.  
The boundary conditions in this work are free-slip (i.e., 
$\omega = 0$) at the top and bottom isobaric surfaces and 
periodic in the zonal~(longitudinal) direction.
Throughout this paper, the lateral coordinates are 
(longitude,\,latitude) $ = (\lambda,\phi)$; 
time, length, pressure, and temperature are expressed 
in units of planetary day ($\tau = 3\!\times\! 10^5$\,s), 
planetary radius ($R_p = 10^8$\,m), reference pressure 
($p_\text{\tiny ref} = 10^5$\,Pa), and reference 
temperature ($T_\text{\tiny ref} = 1500$\,K), 
respectively.  

The $\zeta$--$\delta$--$\Theta$ formulation of the PE in 
equations~(\ref{eq:PE}) facilitates the use of pseudospectral 
method to solve the equations accurately. 
Unlike other numerical methods which offer algebraic convergence
(e.g., finite difference or finite element methods), the 
spectral method offers exponential convergence---i.e., the error 
decays exponentially fast with increased resolution: this leads 
to dramatically improved accuracy for the same or similar 
computational cost \citep[e.g.,][]{Boyd00}. 
In applying the spectral method to solve equations~(\ref{eq:PE}), 
the mapping $\bfv \mapsto \bfv\cos\phi\,$ is employed because the 
components of $\bfv$ are not well suited for representation in 
scalar spectral expansions \citep{Robert66}. 
The formulation of PE in $p$ vertical coordinates also offers a 
practical simplifications of the equations as well as 
clarity of presentation; a second-order finite difference 
scheme is used for the $p$ direction.  

The resolution of the numerical simulations presented here is 
T341L20---i.e., 341 total and zonal wavenumbers ($n$ and $m$, 
respectively) each in the Legendre expansion of the variables 
and 20 vertical levels (layers), spaced linearly in 
$p \in [0.1,1.0]$.
Note that the simulations here may not be numerically converged 
to those that span a much larger $p$-range, especially if the 
density of layers is much greater than the simulations here 
\citep{SkinCho21}.\footnote{A recent study \citep{Menou2020} 
has suggested that high resolution is not necessary for 
hot-Jupiter simulations, but the resolution and dissipation 
order in that study---i.e., $\lesssim$ T682 and 
$\mathfrak{p} < 8$, respectively---are not adequate for 
assessing convergence \citep[see][]{SkinCho20}.} 
In any case, convergence is not the focus here: our focus is on 
a fundamental feature stemming from nonlinearity---acute 
sensitivity to small-scale structures. 
We have verified that the current resolution is adequate to 
lucidly and robustly demonstrate the highlighted feature.
While a variety of $p$ ranges have been used in past simulation 
studies \cite[see, e.g.,][]{Choetal03, showman_2008, RausMen10, TanKoma19, Mend20, SkinCho21}, the range here is chosen to cover majority of 
the thermally irradiated levels while permitting the highlighted 
dynamics effect to be demonstrated clearly.

For the time integration, the second-order leapfrog scheme 
is used with timestep size of $\Delta t = 4.0 \times 10^{-5}$. 
The Robert--Asselin filter \citep{Robert66,Asselin72} with 
a small filter value of $\epsilon = 0.02$ \citep{ThraCho11} 
is applied to suppress the growth of the computational mode 
arising from using the leapfrog scheme to integrate first-order 
(in time) equations. 
All simulations are initialized from rest (i.e., 
$\mathbf{v} = 0$) and evolve under the prescribed thermal 
forcing $\dot{\cal Q}(\bfx,t;\,\tau_r)$ \citep[see, e.g.,]
[Fig.\,1]{SkinCho21}. 
Equations~(\ref{eq:PE}) are integrated to $t = 1000$, 
much longer than the significant dynamical timescales 
(e.g., advective, forcing, and large-scale dissipation 
timescales). 

The only parameters varied in the simulations presented here 
are $\mathfrak{p}$ and $\nu_{2\mathfrak{p}}$ in the 
hyperdissipations, 
\begin{equation}\label{eq:dissip}
  {\cal D}_\chi(\nu,\mathfrak{p})\ =\
  \nu_{2\mathfrak{p}}\!\left[(-1)^{\mathfrak{p}+1}
    \nabla_p^{2\mathfrak{p}}\, +\, 
    \mathcal{C} \right]\,\chi \, , 
\end{equation}
where $\chi \in \{\zeta, \delta, \Theta\}$ and $\mathcal{C}$ 
is a correction term that compensates the damping of uniform 
rotation \citep[][]{Polietal14}. 
Here $\nu_{2\mathfrak{p}}$ of $5.9 \times 10^{-6}$ and 
$1.5 \times 10^{-43}$ (in units of 
$R_p^{2\mathfrak{p}}\,\tau^{-1}$) are carefully chosen for 
$\mathfrak{p}$ of 1 and 8, respectively, to ensure that the 
energy dissipation rate at the truncation wavenumber~$n_t$ 
($=\! 341$) is the same for both $\mathfrak{p}$ values.
At T341 resolution, decreasing $\mathfrak{p}$ and/or increasing
$\nu_{2\mathfrak{p}}$ serve to modulate the energy dissipation 
behavior in small-scale flow structures. 
No other parameterizations (e.g., radiative transfer and 
chemistry) and dissipations or drags (e.g., gravity wave and 
ion) are used; 
currently, these are poorly known for all hot exoplanets and 
their inclusion does not obviate the issue addressed here. The focus of the present study is to 
investigate the dynamics of well-resolved flow structures 
that arise under a large and constant day-night temperature 
contrast.\\

\section{Results}\label{sec:results}

\begin{figure*}
  \vspace*{0.3cm}
  \centering
  \includegraphics[width=\textwidth]{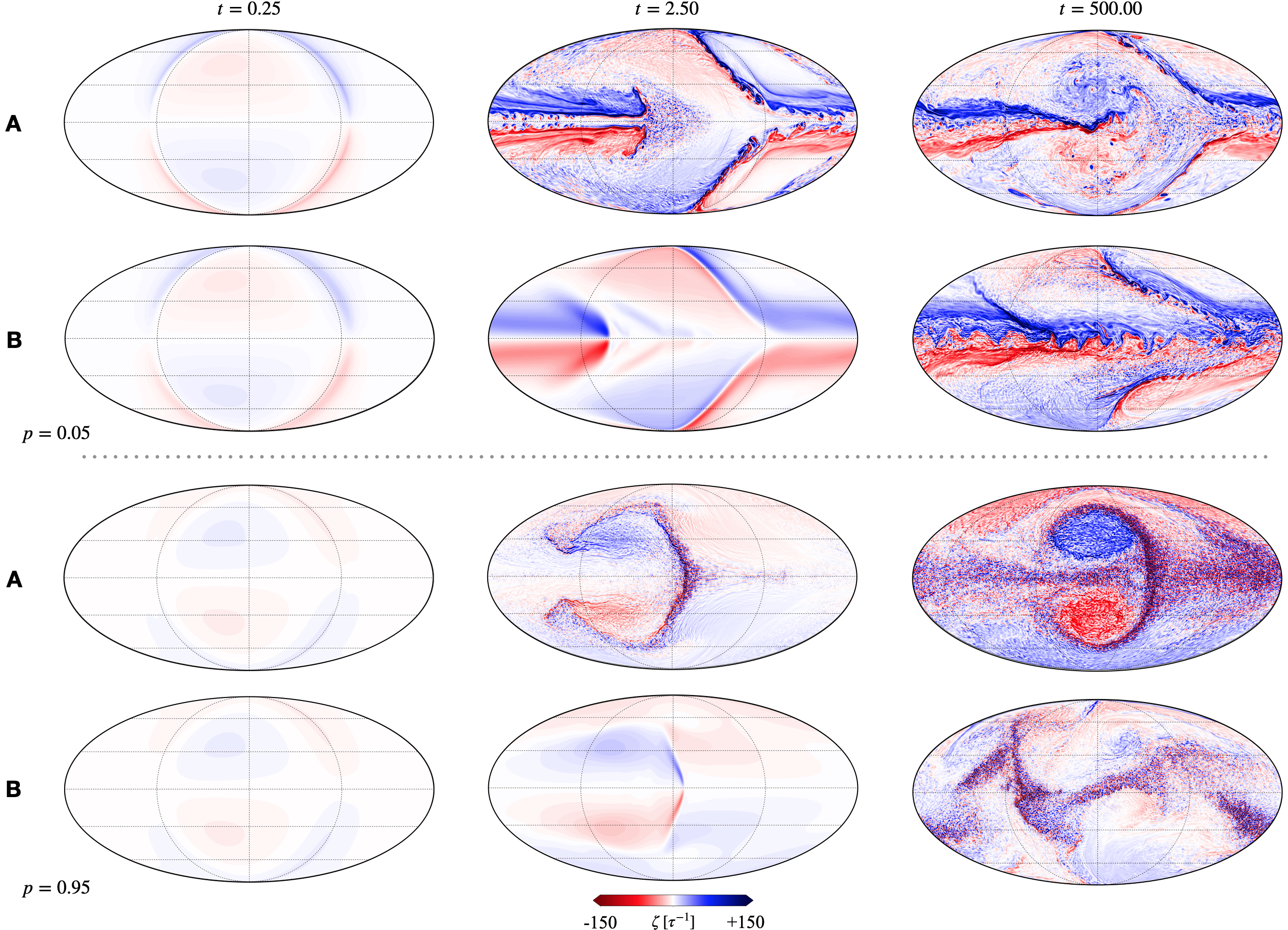}
  \caption{Vorticity field $\zeta$ (in units of $\tau^{-1}$) 
  from two T341L20 resolution simulations ({\bf A} and 
  {\bf B}) at two $p$-levels and three times since $t = 0$. 
  The fields are in Mollweide projection, centered on the 
  substellar point $(\lambda = 0,\, \phi = 0)$; here 
  $\lambda$ is the longitude and $\phi$ is the latitude. 
  Simulations ({\bf A} and {\bf B}) are identical---except  
  {\bf B} uses $\mathfrak{p} = 1$ (i.e., $\nabla^{2}$) 
  dissipation with coefficient $\nu = 5.9 \times 10^{-6}$ 
  (in units of $R_p^2\,\tau^{-1}$), to damp small-scale flow 
  structures more rapidly for $t < 3$; both simulations 
  use the same $\mathfrak{p}$ and $\nu$ thereafter.
  Simulation {\bf A} is a reference simulation, which uses 
  $\mathfrak{p} = 8$ (i.e., $\nabla^{16}$) dissipation with  
  $\nu = 1.5 \times 10^{-43}$ (in units of 
  $R_p^{16}\,\tau^{-1}$), to permit small-scale flow 
  structures to evolve much less encumbered for the entire 
  duration of the simulation ($t = 1000$).
  At $t = 0.25$, the fields from the two simulations are 
  essentially identical at both of the $p$-levels.
  However, at $t = 2.5$, the impact of the difference in 
  damping treatment is clear: numerous small-scale vortices 
  along the fronts, jet flanks, and storm peripheries are 
  entirely missing in simulation {\bf B}. 
  At $t = 500$, the two simulations exhibit significant, 
  qualitative differences---long after the difference in 
  dissipation has ceased; note, e.g., the absence 
  of a strong giant modon in {\bf B}.
  A brief, ``minor'' difference at the small scales very 
  early in the simulation has a persistent, major 
  consequence at the large scale. 
  For more extensive visualizations, including movies, 
  see \citet{SkinCho20,SkinCho21}, \citet{Skinetal23}, 
  and \citet{Changeat_2024}. }
\label{fig:fig1}
\end{figure*} 

Fig.~\ref{fig:fig1} presents the main result of this paper.
When forced by a large day--night temperature contrast 
ramped up on a short timescale, {\it energetic small-scale 
structures quickly emerge in hot-exoplanet atmospheric flows, 
and the preclusion or mitigation of these structures cause 
significant differences in the long-term flow and temperature
distributions}.\footnote{The dynamical state that leads to the 
generation of small-scale structures, such as fast gravity 
waves, is known as an {\it unbalanced} state in geophysical 
fluid dynamics \citep[e.g.,][]{Phillips63,Eliassen84}.}
Here by ``short'' we mean a period smaller than $\sim\! 1$, 
and by ``small'' we mean a lateral size smaller than 
$\sim\! 1/10$.
The significant role of small-scale structures on the flow 
has been noted and addressed from the inception of 
hot-exoplanet atmosphere studies by \citet{Choetal03}, and 
explicitly demonstrated to depend on viscosity and 
resolution in numerical simulations in subsequent 
studies \citep[e.g.,][]{ThraCho11,Choetal15,SkinCho20,Skinetal23}. 
In this paper, we highlight the importance of elongated, 
sharp fronts (that subsequently roll up into long-lived 
vortices) and internal gravity waves \citep{WatCho10}---both 
of which unavoidably arise at the beginning of simulation 
(as well as throughout the simulation): the structures are 
generated in response to the atmosphere's attempt to adjust 
to the applied thermal forcing. 

The figure shows the $\zeta(\lambda,\phi)$ fields from two 
simulations ({\bf A} and {\bf B}) at illustrative $p$-levels 
(0.05 and 0.95) and times (0.25, 2.50, and 500).
The fields are shown in the Mollweide projection, centered at 
the planet's substellar point 
($\lambda\!  =\! 0$,\! $\phi\! =\! 0$).
The two simulations are identical in every way---except 
energy is removed more rapidly in a slightly wider range of 
small scales in simulation {\bf B} than in simulation {\bf A} 
for a very brief interval of time, $t \in [0,3)$.
Two $p$-levels corresponding to near the top and near the 
bottom of the simulation are presented, but the features 
highlighted are generic to other $p$-levels in the 
simulation. 

\begin{figure*}
  \centering
  \includegraphics[width=\textwidth]{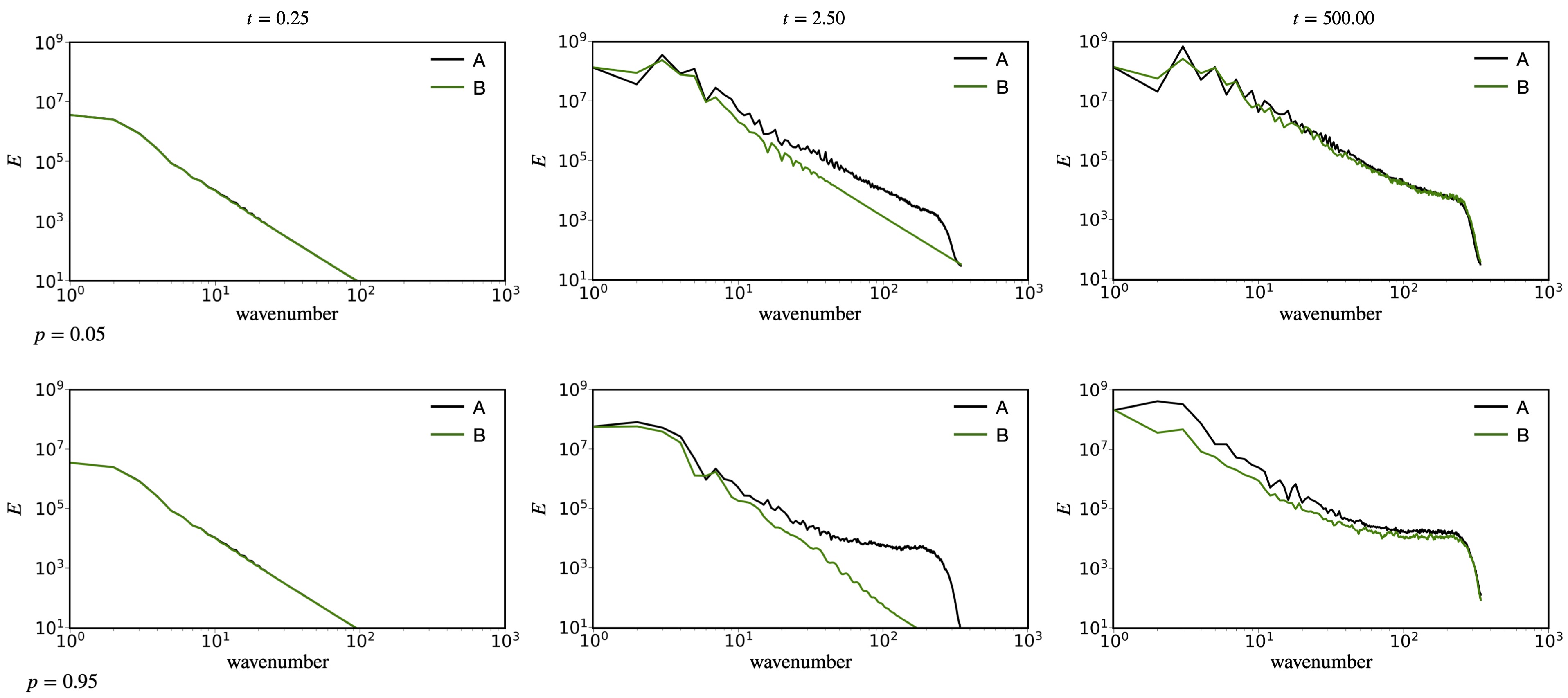}
  \caption{Specific kinetic energy spectrum, $E = E(n)$, 
  of the flows from simulations {\sf A} and {\sf B} in 
  Fig.~\ref{fig:fig1} at $p = 0.05$ (top row) and $p = 0.95$ 
  (bottom row). 
  At $t = 0.25$, the spectra are identical for the two 
  simulations at both $p$-levels. 
  At $t = 2.5$, the spectra at both $p$-levels are 
  markedly different, especially at large $n$, where the 
  spectra for simulation {\sf B} are much steeper than those 
  for simulation {\sf A}.
  This reflects the absence of small-scale vortices along 
  the fronts, jet flanks and vortex peripheries, in 
  simulation {\sf B}. 
  At $t = 500$, long after all the parameters in both 
  simulations have been rendered identical (at $t = 3$), 
  the spectra at both $p$-levels are still very 
  different---particularly at $p = 0.95$; 
  notice the very large deficit of $E$ at small $n$. 
  This is due to the absence of a strong giant modon in 
  simulation {\sf B}.
  The initial difference in the small scales has spread to 
  the large scales, due to the nonlinearity intrinsic in 
  the solved equations. }
\label{fig:fig2}
\end{figure*} 

At $t = 0.25$, the flows of the two simulations are 
essentially identical (cf., {\bf A} and {\bf B} in the 
left column at both $p$-levels).
At this time, the added dissipation in {\bf B} has not had 
a chance to act on the flow (as quantified below).
However, at $t = 2.5$, the difference in dissipation is 
clearly felt by the flow.
For example, sharp vorticity fronts (shear layers) in the 
eastern hemisphere of the dayside and near the equator are 
markedly different (cf., {\bf A} and~{\bf B} in the center 
column at both $p$-levels).
In general, sharp fronts demarcate the outer boundaries 
of planetary-scale hetons\footnote{A heton is a columnar 
vortical structure with opposite signs of vorticity at the 
top and bottom of the column~\cite[e.g.,][]{Kiz06}; 
see {\bf B} in the center column of the figure at the two 
$p$-levels (the hetons are tilted vertically in the eastward 
direction).
Here there are actually four hetons, which comprise two 
modons \citep[][]{HogStom85}---vortex couplets---that spread 
across the equator in each of the $p$-levels.} 
and the flanks of an {\it a}zonal ``equatorial jet''; 
however, unlike in {\bf B}, the fronts have spawned a large 
number of small-scale vortices~(storms) in {\bf A}. 
At $t = 500$, long after the simulations have been brought 
back to the common dissipation condition, the flows are 
still different---and even more so, compared with the flows
at $t = 2.5$.
Hetons are no longer present.
Instead, a cyclonic modon\footnote{Modons composed of two 
cyclones, one in the northern hemisphere with positive vorticity 
and one in the southern hemisphere with negative vorticity; 
unlike hetons, these columnar modons have the same sign of 
vorticity at the top and bottom.} (e.g.,  at the center of 
the frame in {\bf A} at $p = 0.95$) has grown more intense,
while a modon is not present in {\bf B}.
Note that the difference in the flow is not due to a 
temporary ``phase offset'':
the difference persists over the entire duration of the 
simulations, after $t = 3$.
We emphasize here that this difference cannot be accurately 
captured below the T341 resolution because the flow 
structures and their motions are not accurately captured in 
hot, synchronized-planet simulations starting from rest 
\citep{SkinCho20}.

Fig.~\ref{fig:fig2} shows the (specific) kinetic energy 
spectrum, $E = E(n)$, of the flows presented in 
Fig.~\ref{fig:fig1}.
In Fig.~\ref{fig:fig2}, uniform ranges of $E$ and $n$ are 
shown for ease of comparison.
The top row contains the spectra of the flow from the 
$p = 0.05$ level, and the bottom row contains the spectra 
of the flow from the $p = 0.95$ level.
In sum, the figure shows that {\it the difference in 
viscosity, which is limited to the small scales and for 
only a brief period at the beginning of the simulation, 
spreads to large scales and persists in spectral 
space---long after the difference has ceased}. 
The spreading is a fundamental property of nonlinearity 
of equations~(\ref{eq:PE}).  
It also occurs in the full Navier--Stokes equations 
\citep[e.g.,][]{Dobbs-Dixon_2008,Maynetal14,Mendetal16}, 
from which equations~(\ref{eq:PE}) derive.

At $t = 0.25$, the spectra for {\bf A} and {\bf B} are 
identical at both $p$-levels, as expected from the 
corresponding physical space fields in Fig.~\ref{fig:fig1}.
Clearly, the dynamics is not affected by the difference 
in viscosity at this time---{\it at all scales}.
In contrast, at $t = 2.5$, a large difference can be seen 
between the spectra for {\bf A} and {\bf B}---particularly 
at the small scales, $n \gtrsim 10$, as expected.
The difference is huge in the $20 \lesssim n \lesssim 300$ 
subrange.  
At this time all four spectra are still evolving, but the 
overall shape of each one is nearly stationary after 
$t \approx 20$.
Long after the dissipation rate has been rendered identical 
across the entire spectrum (at $t = 3$), the spectra at 
$t = 500$ are still noticeably different---this time much 
more at the large scales ($n \lesssim 10$), especially at 
$p = 0.95$:
at $p = 0.05$, the difference at the large scales is 
significant for only select wavenumbers (e.g., $n = 2$ and 
$n = 3$), but the difference is significant for the entire 
$n \lesssim 20$ subrange at $p = 0.95$.
In fact, at $p = 0.95$, the difference is significant across 
essentially the entire range of well-resolved scales above 
the dissipation range (i.e., $n \lesssim 200$); 
this is again consistent with the corresponding physical 
space fields in Fig.~\ref{fig:fig1}. 

Broadly, energy is accumulated in both the large-scale and 
small-scale subranges (heuristically defined here as 
$n \lesssim 10$ and $n \gtrsim 100$, 
respectively).\footnote{This is unlike in incompressible 
(or, equivalently, small Mach number), homogeneous, 3D 
and two-dimensional (2D) turbulence. 
In the 3D case, energy cascades forward to large~$n$;
in the 2D case, energy cascades backward to small~$n$.}
However, the accumulations are different in the two 
simulations.
As the simulations evolve, their spectra become increasingly 
dissimilar for $n \lesssim 10$, until each simulation reaches 
a different quasi-equilibrium state.
Note that these are the scales which are directly comparable 
with observations as well as explicitly represented in most 
current numerical models.
However, because of the nonlinear interaction, inclusion 
of $n \gg 10$ in the simulation is necessary to accurately 
represent $n \lesssim 10$ \citep[e.g.,][]{Boyd00,SkinCho20}.

Given the general behavior seen here, it is not difficult 
to argue that the difference would only increase with 
resolutions greater than that employed in the present study.  
It also explains in part why hot-exoplanet simulations in 
which small scales have been poorly resolved, or altogether 
missing, would not be able to produce the result of 
high-resolution simulations at the large scales---as pointed 
out in many studies in the past
\citep[e.g.,][]{ThraCho11,PoliCho12,Choetal15,SkinCho20}.  
Because under-resolved and/or over-dissipated models would 
not be able to capture the intrinsic sensitivity and 
complexity of the flow, they would give a specious 
appearance of stability or consistency in their predictions 
for the large scale.

\begin{figure}
  \vspace*{0.3cm}
  \centering
  \includegraphics[width=0.47\textwidth]{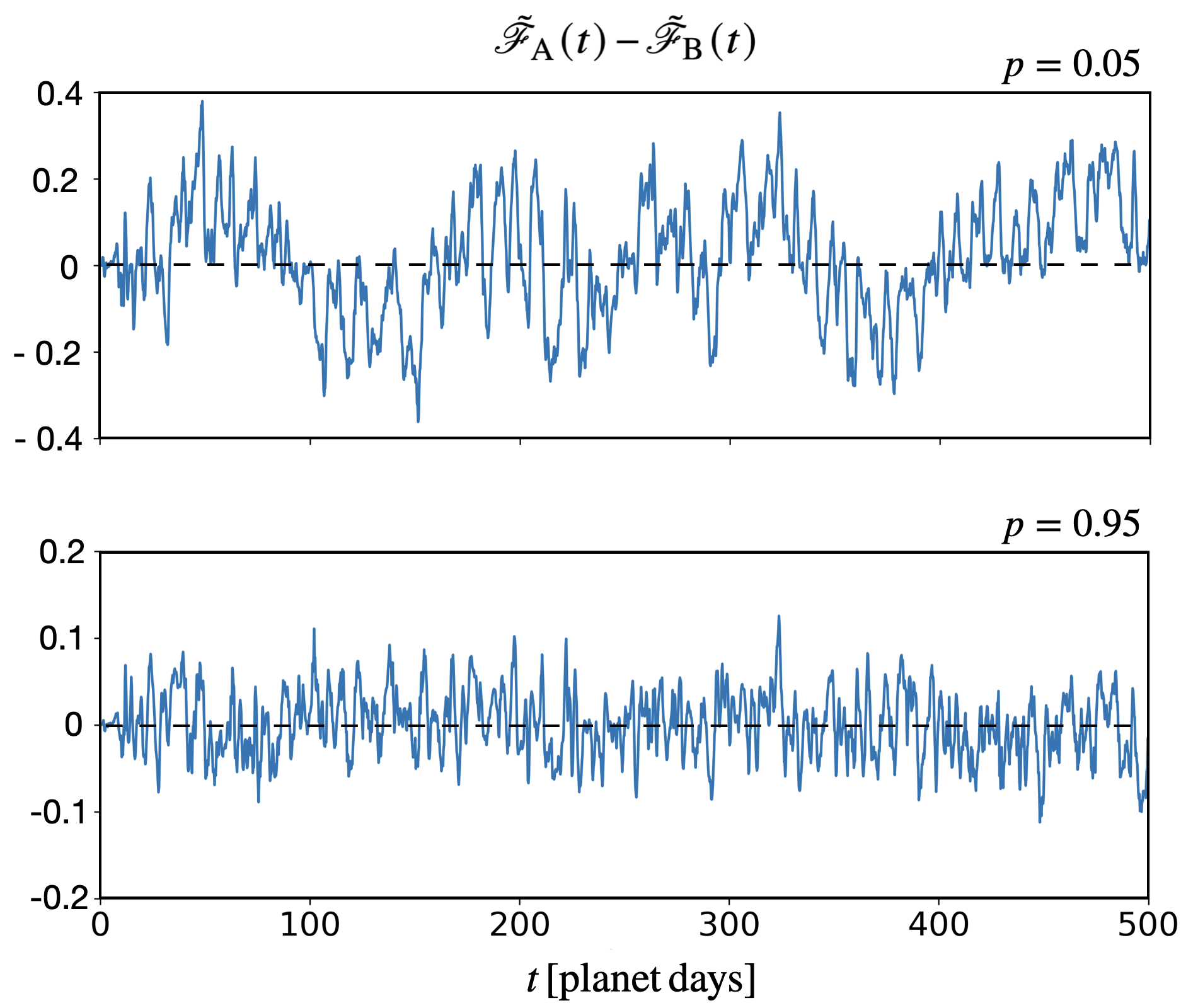}
  \caption{Disk-averaged, normalized, blackbody total emission 
  flux $\tilde{\mathscr{F}}(t)$ {\it difference} between 
  simulations {\bf A} and {\bf B} in Fig.~\ref{fig:fig1}, at 
  the indicated $p$-levels.
  The flux is averaged over a disk centered on the substellar 
  point $(\lambda\! =\! 0,\, \phi\! =\! 0)$, the dayside 
  hemisphere, and weighted by a cosine projection factor 
  (see text). 
  Even under spatial averaging, a significant difference is 
  present---with the difference increasing towards the top of 
  the simulation domain. 
  Here the maximum deviation from the mean corresponds to an 
  averaged temperature ($\langle\,T \rangle_{\mbox{\tiny SS}}$) 
  difference of $\approx\! 100$\,K and $\approx \!75$\,K at 
  $p = 0.05$ and $p = 0.95$ (i.e., $\approx\! 11.5\%$ and 
  $\approx\! 5\%$ of the planet's disk-averaged $T$ at 
  secondary eclipse), respectively.
  The differences are chaotic in time, and hence not due 
  to simple ``phase-shifts'' between the flows in the two 
  simulations.
  Only up to $t = 500$ is shown for clarity, but the 
  behavior is qualitatively the same for up to $t = 1000$, 
  the duration of these simulations. 
  Significantly, the differences are large enough to affect 
  the interpretations of current- and next-generation 
  telescope observations \citep{Rigby_2023,tinetti2021} }.
\label{fig:fig3}
\end{figure}

Fig.~\ref{fig:fig3} presents a clearer picture of the 
behavior over time, particularly for spatially-averaged 
quantities.
As already alluded to, some averaged quantities are directly 
important for observations.
In the figure, the blackbody total emission flux, 
$\mathscr{F}(t) \equiv 
\sigma\langle T^{\, 4}\rangle_{\,\mbox{\tiny SS}}$, is shown 
at two $p$-levels.
Here $\langle\,(\,\cdot\,)\,\rangle_{\,\mbox{\tiny SS\,}}$ 
represents a line-of-sight projection-weighted (a cosine 
factor) average over the dayside disk centered on 
the substellar point~(SS), and $\sigma$ is the 
Stefan--Boltzmann constant; emissivity is assumed 
to be unity, for simplicity.
The flux for each simulation is normalized by the initial
flux, $\tilde{\mathscr{F}}_{0(\cdot)} \equiv
\langle\, T_{0(\cdot)}^{\,4} \rangle_{\,\mbox{\tiny SS}}$,
so that
$\tilde{\mathscr{F}}_{(\cdot)} = 
\tilde{\mathscr{F}}_{(\cdot)} \, /\, 
\tilde{\mathscr{F}}_{0(\cdot)}$;
for example, $\tilde{\mathscr{F}}_{\mbox{\tiny A}} = 
\tilde{\mathscr{F}}_{\mbox{\tiny A}}(t)$ is the normalized 
flux for simulation {\bf A}.
The value of the normalization is same for both simulations
presented and is also independent of the location of the disk 
center, due to the spatially uniform temperature distribution 
used at $t = 0$. 
The $t \in [0,500]$ duration is shown in the figure, but the 
general behavior is unchanged up to $t = 1000$.
We note that an overly long duration (e.g., $t \gtrsim 1000$) 
is unlikely to be physically realistic, given the highly 
idealized initial and forcing conditions employed; 
however, it does illustrate the robustness of the particular 
behavior discussed.

As can be seen, large differences in the averaged flux from 
the simulations persist over a long time and over the entire 
$p$ range.
It is important to stress here again that---because of the 
intrinsic, ever-present imbalance and nonlinearity of the 
atmosphere---the effect of small-scales highlighted is 
{\it not} removed by a simple averaging of 
$\tilde{\mathscr{F}}(t)$ vertically, or in 
time.\footnote{Weighted averaging over $p$ may be desired 
for the purpose of crudely comparing with observed flux 
{\it at a given instant}---if, e.g., the monochromatic 
transfer function \citep{Andrewsetal87} varies greatly over 
the chosen $p$-range.}  
The difference ranges approximately $\pm \, 0.4$ at 
$p = 0.05$ and $\pm \, 0.1$ at $p = 0.95$ for $t\gtrsim 8$, 
and continues for the entire duration of the simulations. 
The variations correspond to disk-averaged temperature 
differences of up to $\pm \, 100$\,K at $p = 0.05$ and 
$\pm \, 75$\,K at $p = 0.95$. 
Such differences, which stem from the acute sensitivities 
inherent in the PE\,\footnote{Note that similar differences 
also arise when different numerical algorithms and/or models 
are employed \citep[see, e.g.,][]{Polietal14,Choetal15}.}, 
directly impact the ability to correctly predict and/or 
interpret observations from current and next-generation 
telescopes such as James Webb Space Telescope and Ariel \citep{Rigby_2023,tinetti2021}.
As noted above, this also underscores the critical 
importance of accurately and consistently representing 
small-scale flows throughout the entirety of the simulation:
the absence of such representation, even for a brief period, 
permanently vitiates the reliability of model predictions.

\begin{figure*}
  \centering
  \includegraphics[width=0.85\textwidth]{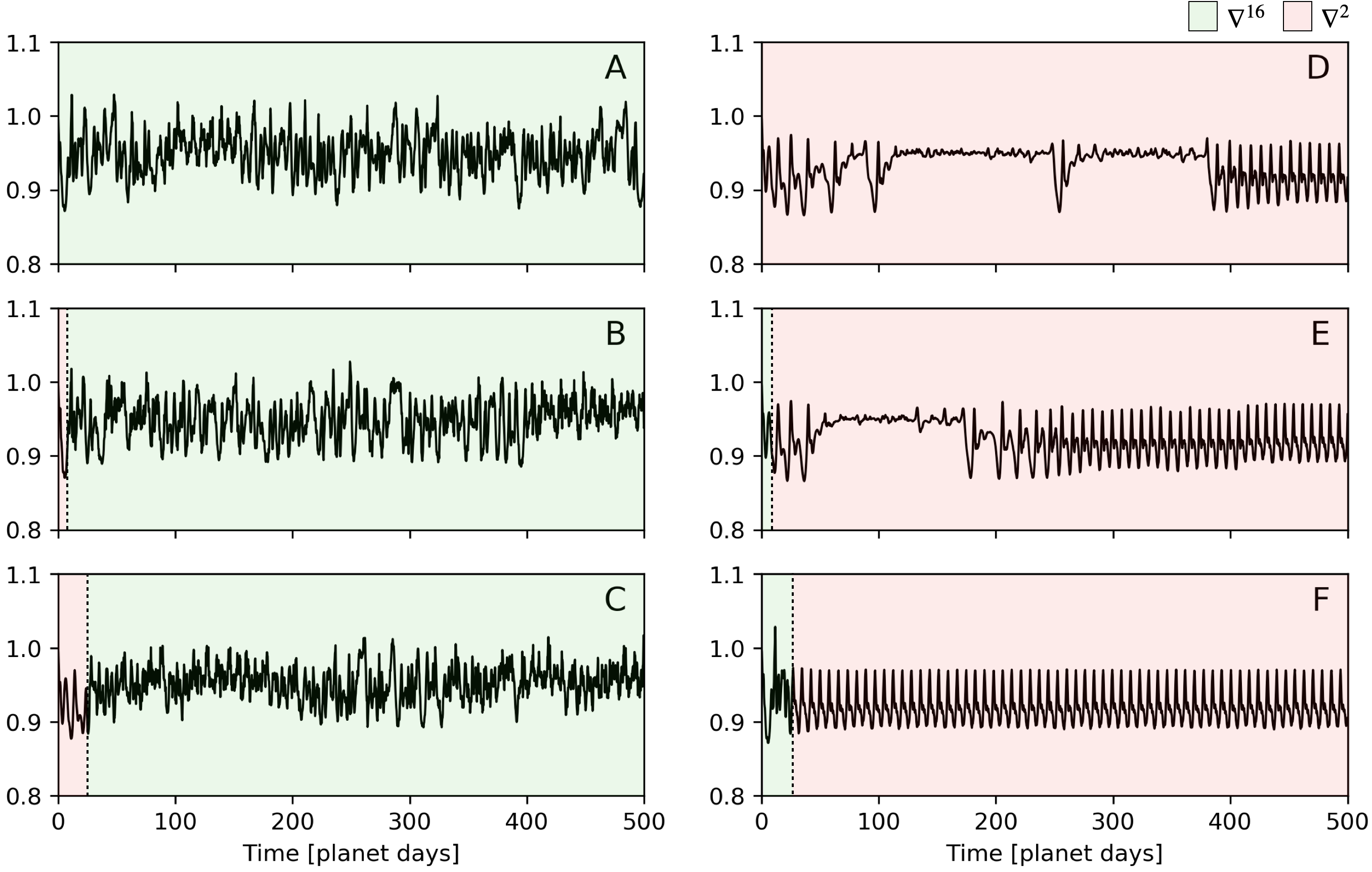}
  \caption{Time-series of $\tilde{\mathscr{F}}(t)$ at 
  $p = 0.95$ for six simulations ({\sf A}--{\sf F}), in which 
  the duration and value of $(\mathfrak{p},\nu)$ pairs, 
  $(8,1.5 \times 10^{-43})$ and $(1,5.9 \times 10^{-6})$, are 
  different. 
  The $\mathfrak{p}$ values, corresponding to $\nabla^{16}$ 
  and $\nabla^2$ operators, are distinguished by the 
  background shading (light green and light red, respectively), 
  and the time at which the dissipation form changes during 
  the simulation is indicated by the dashed line. 
  Otherwise, all simulations are identical---including the 
  values of normalization and $\tilde{\mathscr{F}(0)}$ 
  ($ =\! 1$).
  The simulation in {\sf A} is the simulation in 
  Fig.~\ref{fig:fig1}A. 
  ({\sf A},\! {\sf D}) are reference simulations, with 
  $\mathfrak{p}$ and $\nu$ fixed for the entire duration. 
  ({\sf B},\! {\sf C}) correspond to simulations that are 
  identical to {\sf A}, but with enhanced dissipation of 
  small scales for $t < 3$ and $t < 20$, respectively. 
  ({\sf E},\! {\sf F}) correspond to simulations that are 
  identical to {\sf D}, but with reduced dissipation of 
  small scales for $t < 3$ and $t < 20$, respectively.
  Having the reduced dissipation on for a longer period 
  appears to hasten the transition to a large-amplitude, 
  long-period oscillatory state (cf., {\sf D}--{\sf F});
  and, having the enhanced dissipation on for a longer 
  period appears to introduce a very long-period 
  oscillation (cf., {\sf A}--{\sf C}). 
  Differences in dissipation rate of small scales 
  influence the evolution in complex ways.
  }
\label{fig:fig4}
\end{figure*}

Fig.~\ref{fig:fig4} shows a more complete picture of the 
differences at high resolution. 
The figure presents $\tilde{\mathscr{F}}(t)$ at $p = 0.95$ 
for six T341L20 simulations ({\sf A}--{\sf E})---all 
identical except for the $(\nu,\,\mathfrak{p})$ pairs, 
$(1.5 \times 10^{-43},\,8)$ and $(5.9 \times 10^{-6},\,1)$;
see Eq.~\ref{eq:dissip}. 
In each simulation, the applied dissipation is switched to 
another one from $\{\bfnabla^{16},\bfnabla^2\}$ at the time
indicated by the dashed line in the figure.
Panels~{\sf A} and~{\sf D} present reference simulations, 
with no change in dissipation for all~$t$. 
Panels~{\sf B} and~{\sf C} present simulations with 
parameters identical to those in the simulation of {\sf A} 
but with a stronger dissipation rate, 
$(\nu,\,\mathfrak{p}) = (5.9 \times 10^{-6},\,1)$, 
applied at the beginning for different durations ($t < 3$ 
and $t < 20$, respectively).
Panels~{\sf E} and~{\sf F} present simulations with 
parameters identical to those in the simulation of {\sf D} 
but with a weaker dissipation rate, 
$(\nu,\,\mathfrak{p}) = (1.5 \times 10^{-43},\,8)$, applied 
at the beginning for different durations ($t < 3$ and 
$t < 20$, respectively, here as well).

Broadly, two distinct types of $\tilde{\mathscr{F}}(t)$ 
behaviors emerge \citep[cf.,][]{SkinCho20}: 
1) a dynamic and chaotic large-scale flow characterized 
by multiple, persistent large-amplitude oscillations; 
and, 2) persistent, regular oscillations that ``kick in'' 
after a period of small-amplitude oscillations.
The two types can be seen in the left and right columns of 
Fig.~\ref{fig:fig4}, respectively.
The first type is caused by dynamical instability and 
turbulent motion of energetic, planetary-scale vortices, 
which ultimately migrate around the planet; these giant 
vortices, which may be singlets or doublets 
\citep{Skinetal23}, interact with a large number of small 
vortices during the migration in a way reminiscent of 
Brownian motion (panels~{\sf A}--{\sf C}).  
In this case, the planet's vorticity and temperature 
fields are highly inhomogeneous, with strong meridional 
(north--south) asymmetries and vigorous mixing on the 
large scale. 
The second type is caused by a long-lived, planetary-scale, 
meridionally (latitudinally) symmetric modon, which is weaker 
(lower $|\zeta|$) than the giant vortices in the first type
of behavior.
After a transient period of large excursions from near the 
substellar point at the beginning of the simulation, there 
is generally a period of ``quiescence'', when the modon's 
position is nearly fixed at the substellar point (e.g., 
$120 \lesssim t \lesssim 260$ in panel~{\sf D}).  
After this quiescent period, the modon transitions to a one
of westward ``migration'' around the planet---subsequently 
either transitioning back to the quiescent state 
($t \approx 275$ in panel~{\sf D}) or remaining in the 
migrating state ($t \approx 230$ in panel~{\sf D}). 

Notice that the quiescent state is not always present in 
a simulation (panel~{\sf F}).
However, when it is present (panels~{\sf D} and~{\sf E}), 
there is nearly a fourfold reduction in 
$\tilde{\mathscr{F}}$ variations as well as a sustained 
high amplitude in $\tilde{\mathscr{F}}$ compared with 
$\tilde{\mathscr{F}}$ in the migrating state (present in 
all of panels~{\sf D}--{\sf F}). 
Both the reduction of variation and sustenance of high 
amplitude occur because there is little or no heat 
transport away from the dayside. 
In contrast, when the modon migrates westward, it mixes 
and advects heat to the western terminator or beyond.
The timescale of the mixing/advection is relatively short, 
evinced by the fast decay time of the regular peaks seen 
in panels~{\sf D}--{\sf F}: it is roughly equal to the 
thermal relaxation timescale for the $p$-level shown in 
the figure. 

It is clear from Fig.~\ref{fig:fig4} that model predictions 
for the large scale are highly sensitive to the dissipation 
rate of the small scales. 
It follows that the sensitivity would be entirely missed 
if the small scales are not represented in simulations. 
In the figure, the right column shows that the sensitivity 
is active even when the small scales are heavily suppressed 
throughout the majority of the flow's evolution, as long as 
the small scales are represented; cf., left column.
In addition, the evolutions in {\sf E} and {\sf F} are 
noticeably different, despite the simulations being 
identical except for the small difference in the duration 
of the reduced dissipation at the beginning. 
Less dramatic, but still noticeable, behaviors are seen 
in the opposite situation, in which the small scales are 
more heavily suppressed for the first 3 and 20 days (only) 
of the evolution ({\sf B} and {\sf C}, respectively). 
In {\sf C}, a long-period variation not seen in {\sf A}
appears in the evolution; in {\sf B}, a suggestive
transition to a ``quiescent''-like state is observed 
(cf., {\sf A}). 
Finally, it is important to also note that the temporal 
mean values of  $\tilde{\mathscr{F}}$ (hence 
$\langle\, T\,\rangle_{\mbox{\tiny SS}}$) vary among 
the simulations at quasi-equilibration---even though the 
thermal forcing is identical in all of them.\footnote{We 
remind the reader that ``equilibration'' depends on the 
realism and completeness of the forcing and initial 
conditions supplied to the model, {\it not} on the 
steadiness of (averaged) model outputs.}


\section{Discussion}\label{sec:conclusion}

In this paper, we have presented results from 
high-resolution atmospheric flow simulations with a setup 
(initial, boundary, and forcing conditions) commonly 
employed in hot-exoplanet studies.
Our simulations explicitly demonstrate that the behavior of 
the atmosphere at {\it the large-scale is highly sensitive 
to the rate of energy loss in the small scales}---the loss 
both intentional and not.
Surprisingly, the sensitivity is present even if the 
increase in the rate is operating only for a very brief 
period. 
Hence, deviation from high-resolution results are fully 
expected when the small scales are poorly resolved or 
altogether missing, as have been demonstrated by 
\citet{SkinCho20}.
As in that study, the sensitivity is comprehensively 
illustrated in the physical, spectral, and temporal spaces 
in this study.
Here we clearly demonstrate that high-resolution is 
necessary for generating accurate predictions, as 
the small scales non-trivially affect the evolution of
flow and temperature at the large-scale.

More broadly, high resolution (as well as an accurate algorithm) 
is also critical for understanding ageostrophic (unbalanced) 
turbulence\footnote{which is characteristic of 
hot-exoplanet atmospheres}, in general.
It is found that the presence---or preclusion---of the 
small-scale structures, which {\it appear almost immediately 
in the flow} ($t \lesssim 1$), leads the hot-exoplanet 
atmosphere simulations to settle into qualitatively 
different quasi-equilibrium states.
The small-scale flow structures generated at early times of 
the evolution are {\it i}) sharp, elongated fronts that roll 
up into energetic vortices and {\it ii}) radiated, internal 
gravity waves. 
These form in response to the atmosphere's attempt to 
adjust to the applied forcing---unrelated to the degree to 
which the radiative process is idealized: 
the aggressive response is due to the rapidity of the thermal
relaxation to a high day--night temperature gradient, leading 
to large Rossby and Froude 
numbers \cite[see, e.g.,][]{Choetal08} for the flow. 
The need for high resolution and balancing to address such 
atmospheres has been noted since the beginning of exoplanet 
atmospheric dynamics studies by \cite{Choetal03}: 
in that study, nonlinear balancing and slow lead-up to the
full thermal forcing have been employed at T341 resolution.

This work has significant implications for general circulation 
and climate modeling of hot-exoplanets. 
Models that use any combination of low order dissipation, 
low spatial or temporal resolution, high explicit viscosity 
coefficients, or strong basal drags to force numerical 
stabilization are at risk of generating inaccurate and/or 
unphysical solutions.
This is because all of the above expediencies prevent 
small-scale flows from being adequately captured throughout 
the simulation's evolution. 
In our view, it is unlikely that the state of the 
hot-exoplanet atmosphere can be usefully captured in such 
models---as the dynamics, which forms the backbone on which 
physical parametrizations hang, is itself questionable. 

Accurately simulating exoplanet atmospheres is a very complex 
and difficult problem, requiring meticulous assessment and 
reduction of uncertainty at every level of the model 
hierarchy---from the equations solved, to the dynamical 
core that generate the solutions, to the parameterizations 
that enhance the dynamics as well as rely on it.
This is the case even for the Earth, for which detailed
observation-derived (referred to as ``analyzed'' in numerical 
weather and climate predictions) inputs to the numerical 
models are available {\it and} the dynamics are 
geostrophically balanced (small Rossby and Froude numbers).
Importantly, effects of small-scales very similar to those 
described here are well known in Earth climate simulation 
studies~\citep{Rialetal04, Sriveretal15, Deseretal20}.
However, as expected, their effects are weaker compared to 
those for hot-exoplanets, which are far from geostrophic 
balance. 
Despite this, the effects seem not to have garnered much 
attention in exoplanet studies thus far. 
The preclusion of small-scale structures poses a 
particularly critical problem in hot-exoplanet ``radiative 
transfer/chemistry-coupled dynamics'' simulations.
The high sensitivity of the overall flow to small-scale 
structures, which arise early in the simulation, means 
that considerable care must be taken to: 
{\it i}) accurately represent structures of wide-ranging scales 
throughout the entire duration of the simulation, and
{\it ii}) sensibly initialize the simulation.\\

\section{Acknowledgments}
The authors thank Joonas N{\"a}ttil{\"a} and Quentin 
Changeat for helpful discussions.  
JWS is supported by NASA grant 80NSSC23K0345 and a Simons 
Foundation Pivot Fellowship awarded to Albion Lawrence.
This work used high performance computing at the San Diego 
Super Computing Centre through allocation PHY230189 from 
the Advanced Cyberinfrastructure Coordination Ecosystem: 
Services \& Support (ACCESS) program, which is supported 
by National Science Foundation grants \#2138259, \#2138286, 
\#2138307, \#2137603, and \#2138296.
This work also used high-performance computing awarded by 
the Google Cloud Research Credits program GCP19980904. 

\vspace{5mm}

\bibliography{references}
\bibliographystyle{aasjournal}

\end{document}